\documentclass[
 reprint,twocolumn,
 groupedaddress,
 superscriptaddress,
 amsmath,amssymb,
 aps,
 prx,
]{revtex4-1}

\usepackage{graphicx}% Include figure files
\usepackage{bm}% bold math
\usepackage{bbold}
\usepackage[usenames,dvipsnames]{color} %for color
\usepackage[colorlinks=true , citecolor=blue,urlcolor=blue]{hyperref}

\newcommand{\lr}[1]{\left( #1\right)}

\renewcommand{\figurename}{\textbf{Fig.}}
\begin{document}

% \preprint{APS/123-QED}
% \begin{CJK*}{UTF8}{}
\title{Observation of Floquet prethermalization in dipolar spin chains}% Force line breaks with \\

\author{Pai Peng }\email[]{paipeng@mit.edu}
\affiliation{Department of Electrical Engineering and Computer Science, Massachusetts Institute of Technology, Cambridge, MA 02139}
\author{Chao Yin}\thanks{P.P. and C.Y. contributed equally to this work.}
% \thanks{These authors contributed equally to this work.}
\affiliation{
Research Laboratory of Electronics, Massachusetts Institute of Technology, Cambridge, Massachusetts 02139, USA
}
\author{Xiaoyang Huang}
\affiliation{
Research Laboratory of Electronics, Massachusetts Institute of Technology, Cambridge, Massachusetts 02139, USA
}
\author{Chandrasekhar Ramanathan}
\affiliation{Department of Physics and Astronomy, Dartmouth College, Hanover, NH 03755, USA}
\author{Paola Cappellaro}\email[]{pcappell@mit.edu}
\affiliation{Department of Nuclear Science and Engineering, Massachusetts Institute of Technology, Cambridge, MA 02139}
\affiliation{
Research Laboratory of Electronics, Massachusetts Institute of Technology, Cambridge, Massachusetts 02139, USA
}
\date{\today}
\maketitle

\textbf{
Periodically driven \textit{Floquet} quantum systems provide a promising platform to investigate novel physics out of equilibrium~\cite{Eisert15}. Unfortunately, the drive generically heats up the system to a featureless infinite temperature state~\cite{Lazarides14,DAlessio14,Kim14}. For large driving frequency, the heat absorption rate is predicted to be exponentially small, giving rise to a long-lived prethermal regime which exhibits all the intriguing properties of Floquet systems~\cite{Abanin17,Abanin15,Kuwahara16,Abanin17b}. Here we experimentally observe Floquet prethermalization using nuclear magnetic resonance techniques. We first show the relaxation of a far-from-equilibrium initial state to a long-lived prethermal state, well described by the time-independent ``prethermal'' Hamiltonian. By measuring the autocorrelation of this prethermal Hamiltonian we can further experimentally confirm the predicted exponentially slow heating rate.
More strikingly, we find that in the timescale when the effective Hamiltonian picture breaks down, the Floquet system still possesses other quasi-conservation laws. Our results demonstrate that it is possible to realize robust Floquet engineering, thus enabling the experimental observation of  non-trivial Floquet phases of matter.
}

% \end{CJK*}

% \section{Introduction}
Driving quantum systems out of equilibrium promises to reveal new physics phenomena beyond equilibrium statistics~\cite{Eisert15}. In particular, periodically driven, or Floquet, systems have received great attention:
If observed only stroboscopically, a Floquet system can simulate a time-independent Hamiltonian that might not be otherwise directly accessible, ultimately enabling universal quantum simulation~\cite{Lloyd96,Childs18}. %todo: deleted: Trotter59 via the Trotter-Sukuzi scheme \cite{Trotter59}
Applications range from modifying the tunneling and coupling rates in lattice systems~\cite{Eckardt05,Tsuji11,Gorg18} to inducing non-trivial topological structures~\cite{Lindner11,Wang13s}.
%todo: deleted: Kitagawa10
More strikingly, Floquet systems exhibit novel phenomena that have no static counterparts, including discrete-time crystalline phases~\cite{Choi17n,Zhang17n}
%, dynamical localization~\cite{Dunlap86,Fishman82}
and dynamical phase transitions~\cite{Bastidas12}.

However, a generic interacting Floquet %many-body
system absorbs energy from the drive and is expected to heat up to infinite temperature, thus suppressing all the interesting phenomena~\cite{Lazarides14,DAlessio14,Kim14}.
Many-body~\cite{Lazarides15,Ponte15,Bordia17,Khemani16,Wei18} and dynamic localization~\cite{Heyl19,Sieberer19,Ji18} provide a way to escape thermalization, as well as some fine-tuned driving protocols~\cite{DAlessio13}.
%todo: deleted: Prosen98
More generally, it has been theoretically shown~\cite{Abanin17,Abanin15,Kuwahara16,Abanin17b} that the dynamics of a Floquet system under rapid drive (fast compared to any local energy scale) is approximately governed by a time-independent effective Hamiltonian (called ``prethermal Hamiltonian'') up to a correction exponentially small in the driving frequency.
This property is generic to any system with local interactions, without requiring disordered fields or fine-tuned parameters.
The Floquet system can be approximately described by the prethermal Hamiltonian for an exponentially long time, with emergent symmetries or ``quasiconserved quantities'' (i.e., conserved by the prethermal Hamiltonian but not by the exact Floquet propagator)~\cite{Yin2020}, such as (prethermal) energy conservation~\cite{Abanin17} and Ising symmetry~\cite{Else17}.
These quasiconserved quantities demonstrate extraordinary robustness in quantum simulation~\cite{Heyl19,DAlessio13,Sieberer19}, and the emergent symmetries set the foundation of Floquet phases~\cite{Else17,Luitz19,Machado2020}.

 Floquet prethermalization, featuring exponentially slow heating, has been confirmed numerically in several Hamiltonian models~\cite{Herrmann18,Bukov16b,Luitz19,Mallayya19}, but an experimental study is still missing.
In this paper, we experimentally observe Floquet prethermalization in a natural nuclear spin system by developing nuclear magnetic resonance (NMR) techniques to tune the driving frequency, while keeping experimental errors constant.
 Intriguingly, we find a quasiconserved observable with an even slower heating rate than the prethermal energy, indicating that emergent symmetries and associated Floquet phases may exist beyond the effective Hamiltonian picture.

\begin{figure*}[!htp]
\centering
\includegraphics[width=1.0\textwidth]{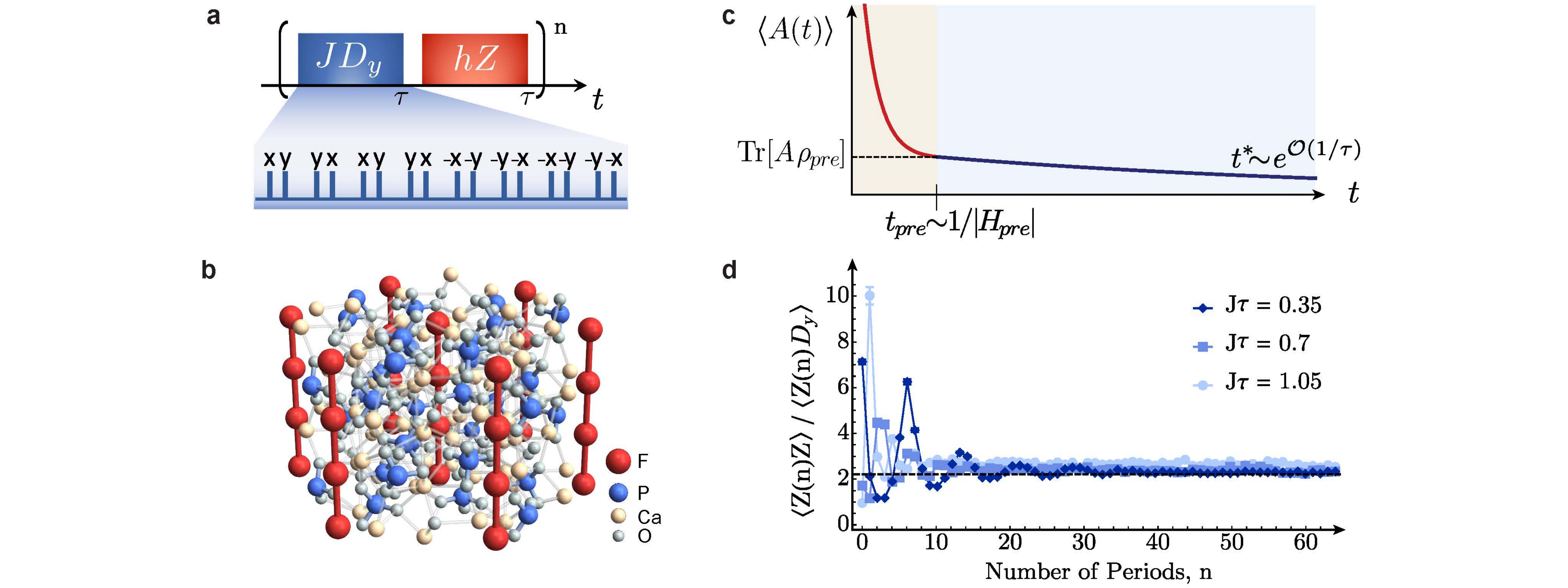}
\caption{\label{ZZ_ZDy}
\textbf{Schematic of the experimental system and Floquet prethermalization.}
(a) Floquet driving scheme for the kicked dipolar model. In the inset: 16 RF $\pi/2$-pulse sequence~\cite{Wei18,Wei19} used to engineer the $D_y$ dipolar Hamiltonian up to second order in the Magnus expansion, with variable strength $J$ and fixed time and control errors (\textsf{x,y} indicate the pulse phases). The variable-strength kicking field, $hZ$, can be introduced by phase-shifting all pulses by an angle $h\tau$ without physically applying a field~\cite{Wei19}, an extremely robust experimental method. In the following we fix $h/J=1$.
(b) Fluorapatite crystal, Ca$_5$(PO$_4$)$_3$F: the $^{19}$F spins-1/2 (red) form linear chains.
(c)
Cartoon showing the typical thermalization process of a generic observable $A$ in the fast driven Floquet systems: the observable has an initial fast decay (orange shaded area) to its quasiconserved, prethermal value, followed by a slow relaxation toward the fully thermalized value (blue shaded area).
(d) Prethermalization of the $Z$ magnetization, $\langle Z(n)Z\rangle/\langle Z(n)D_y \rangle$ in the kicked dipolar model as a function of $n$ for different $J\tau$. The black dashed line is the ratio
$\langle \overline H Z\rangle/\langle \overline H D_y \rangle=8/[3\zeta(3)]\approx2.2$.
Error bars are determined from the noise in the free induction decay~\cite{SOM}.}
\end{figure*}

%%%%%
% To do: Paola: should we switch fig. 1(a) and (b)
Experiments are conducted on nuclear spins in fluorapatite, %~\cite{VanderLugt64}
an experimental system recently used to show many-body localization~\cite{Wei18} and static prethermalization~\cite{Wei19}.
The system can be modeled as an ensemble of chains of $^{19}$F spins-1/2 $S^j$ [Fig.~\ref{ZZ_ZDy}(b)] interacting (see methods) via the dipolar Hamiltonian $H_\text{Dipz}=J_0 D_z$,
%+\sum_j h_j S_z^j$,
where $D_\alpha=\sum_{j<k}^L \frac{1}{2|k-j|^{3}}\left(3S_\alpha^j S_\alpha^k - \vec{S}^j\!\cdot\!\vec{S}^k \right)$. The large system size ($L>50$) is crucial to studying thermalization, which only happens in the thermodynamic limit.

The initial state at room temperature (see methods) is given by $\rho_0\!\approx\!(\mathbb{1}\!-\!\epsilon Z)/2^L$, with $\delta\rho_0\equiv Z=\sum_j S_z^j$ the collective z-magnetization, which gives rise to the signal. Importantly, the signal can be rewritten as the two-point correlation at infinite temperature, $\text{Tr}[\delta\rho(t)\mathcal{O}]/2^{L}\equiv \langle \delta\rho(t)\mathcal{O}\rangle_{\beta=0}$, where $\delta\rho$ also plays the role of an observable.
Using RF controls, we can engineer $\delta \rho(0)$ and $\mathcal{O}$ to be not only collective magnetization around any axis, but also (see methods) the dipolar order operator $D_\alpha$ in any arbitrary direction $\alpha$.

To probe Floquet prethermalization, here we  consider the kicked dipolar model, with Floquet propagator $U_F=e^{-ihZ\tau}e^{-iJD_y\tau}$ [Fig.~\ref{ZZ_ZDy}~(a)].
$J$ and $h$ are the strength of the dipolar interaction and the collective z-field respectively. While inspired by the kicked Ising model, as we will show this model presents a richer prethermalization structure.
%As we will show, this model presents two quasiconserved quantities: prethermal energy and dressed dipolar order.
Typically $J$ is fixed by the system properties (here the crystal lattice), and probing Floquet thermalization requires varying the time $\tau$, introducing undesired effects from, e.g., decoherence.
To overcome this issue, in our experiments we engineer the Hamiltonian $JD_{y}$ by modulating~\cite{SOM} the natural dipolar Hamiltonian $H_\text{Dipz}=J_0D_z$ with $n$ repetitions of the pulse sequence in Fig.~\ref{ZZ_ZDy}~(a).
Then, to investigate different driving frequencies, we vary the effective dipolar strength $J$ by changing only relative delays within the sequence, while keeping the sequence length fixed, $\tau=120~\mu$s. The advantage is that the total experimental time $n\tau$ and number of pulses $16n$ are kept constant for different driving frequencies, $J\tau$. As a result, the effects of decoherence and control errors do not change when varying $J\tau$~\cite{SOM}. This allows us to isolate Floquet heating due to the coherent quantum evolution from the presence of experimental imperfections.

When the driving frequency is large compared to local energy scales, the Floquet dynamics is captured by a time-independent prethermal Hamiltonian $H_{pre}$, obtained from the truncated Floquet-Magnus expansion,
plus an exponentially small~\cite{Kuwahara16, Abanin17} time-dependent term, $\|\delta H(t)\|<\exp\left(-O(1/\tau)\right)$ (see methods).
The typical thermalization process for an observable $A(t)$ is shown in Fig.~\ref{ZZ_ZDy}(c). In a time $t_{pre}\!\sim\!1/\|H_{pre}\|$, $\langle A(t)\rangle$ prethermalizes to its canonical ensemble value $\mathrm{Tr}\left(A\rho_{pre}\right)$ set by the Gibbs state of the prethermal Hamiltonian,
$\rho_{pre}=e^{-\beta H_{pre}}/\mathcal{Z}$,
with $\mathcal{Z}$ the partition function
and $\beta$ determined by the initial state energy. After this transient, $\langle A(t)\rangle$ decays under the influence of $\delta H$, and the effective Hamiltonian picture gradually breaks down. Finally, the system thermalizes to infinite temperature in a timescale $t^*\!\sim\!\exp(O(1/\tau))$.

\begin{figure*}[!htp]
\centering
\includegraphics[width=1.0\textwidth]{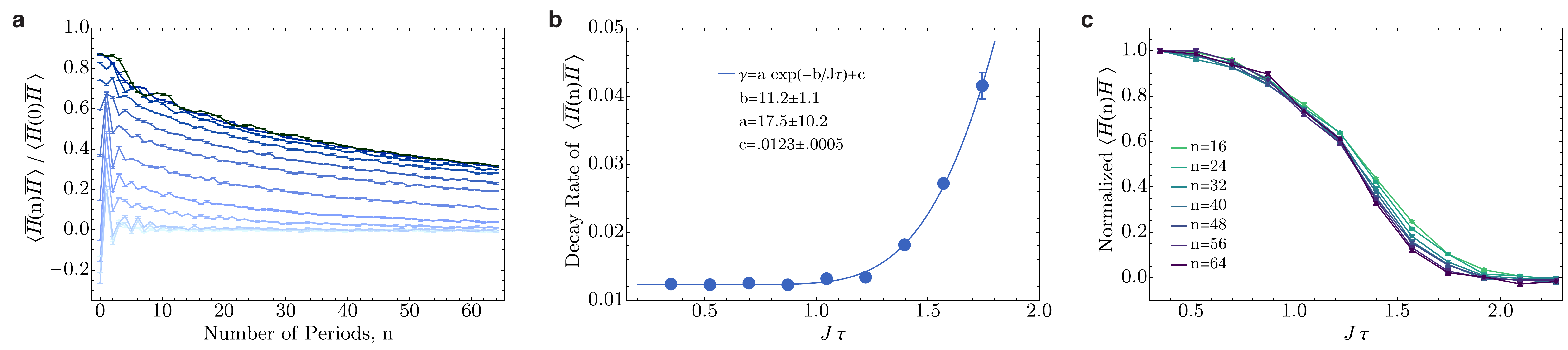}
\caption{\label{Hid}
\textbf{Breakdown of energy conservation.}
(a) Autocorrelation as a function of $n$, for $J\tau=0.35$ to 2.27 in steps of 0.175. Darker to lighter colors represent smaller to larger $J\tau$.
(b) We fit the autocorrelations from $n=20$ to $n=64$ to an exponentially decaying function $\exp(-\gamma n)$ and plot the decay rate $\gamma$. The length of the error bars corresponds to two standard deviation of the fitted decay rate. The solid curve is a fit to $\gamma=a\exp(-b/J\tau)+c$.
(c) Autocorrelation versus $J\tau$ for different $n$. Lighter colors represent smaller $n$ and darker colors represent larger $n$. For each $n$, the autocorrelation is normalized by its value at $J\tau=0.35$ , i.e. the leftmost point is normalized to 1. In (a) and (c), error bars are determined from the noise in the NMR free induction decay~\cite{SOM}.
}
\end{figure*}

To experimentally demonstrate prethermalization, we consider the dynamics of $\delta\rho(n)\!\propto\!Z(n)$ in the kicked dipolar model, where for any observable
$\mathcal{O}(n)=\left(U_F\right)^n \mathcal{O}(U_F^\dagger)^n$.  %under Hamiltonian in a time $t<\exp(O(1/\tau))$.
%For KDM, the average Hamiltonian is $\overline H^\mathrm{(K)}=Z+D_y$.
We plot the ratio of the experimentally measured two-point correlators $\mathrm{Tr}[\delta\rho(n)Z]$ and $\mathrm{Tr}[\delta\rho(n)D_y]$ in Fig.~\ref{ZZ_ZDy}(d).
We take the ratio because it is insensitive to experimental imperfections~(see~\cite{SOM}).
% and it is independent of $\epsilon'$.
For fast driving, $J\tau=0.35$, the ratio quickly stabilizes to a non-zero value after a few oscillations, indicating the system reached a quasi-equilibrium.
%This prethermal equilibrium state corresponds to the high-temperature Gibbs state of the prethermal Hamiltonian $(\mathbb{1}/2^L\!-\!\rho_{pre})\!\approx\!\epsilon'H_{pre}$, which is well approximated by the first-order average Hamiltonian, $\overline H=JD_y+hZ$.
Indeed, the initial density matrix %$\rho_0=\mathbb{1}-\epsilon\delta\rho$
prethermalizes to the the high-temperature Gibbs state of the prethermal Hamiltonian $\rho_{pre}=\mathbb{1}-\epsilon' H_{pre}\approx\mathbb{1}-\epsilon' \overline{H} +O(\tau)$, where $\overline{H}=JD_y+hZ$ is the zeroth-order average Hamiltonian (see methods).  As a result, the ratio saturates at $\langle \overline HZ\rangle/\langle \overline HD_y \rangle=8/[3\zeta(3)]$  (with $\zeta(3)\equiv\sum_{n=1}^\infty n^{-3}$ the Riemann zeta function). For slightly slower driving, the ratio still stabilizes, but its long-time value deviates from $8/3\zeta(3)$ due to the presence of higher order terms in $H_{pre}$.
The slow decay to zero induced by the error term $\delta H$ is not evident in the ratio shown in Fig.~\ref{ZZ_ZDy}(d). To see exponentially slow heating we need to look at observables conserved by $H_{pre}$.

An obvious conserved quantity is the Hamiltonian $H_{pre}$ itself: prethermal energy quasiconservation can naturally reveal the prethermal phase, and its breakdown indicates the eventual heating to infinite temperature.
%
% \subsection{Breakdown of energy conservation}
In experiments, we can only measure the average Hamiltonians $\overline H$, which still serves as a good approximation to $H_{pre}$ even at long times. Indeed, during a short transient of order $t_{pre}$, terms in $\overline H$ not in $H_{pre}$ prethermalize (creating highly correlated operators that cannot be observed). After this prethermalization process, $\langle \overline{H}(n)\overline{H}\rangle$ and $\langle H_{pre}(n)H_{pre}\rangle$ differ only by a constant factor, and they both undergo a slow decay.
We obtain the autocorrelation of $\overline H$ by adding 4 experiments $\langle Z(n)Z\rangle$, $\langle Z(n)D_y\rangle$, $\langle D_y(n)Z\rangle$, and $\langle D_y(n)D_y\rangle$, as shown in Fig. \ref{Hid}(a). The initial damped oscillations signal the prethermalization of $\overline H(n)$ under $H_{pre}$. This prethermalization stage is followed by a slow exponential decay as a result of heating.
We plot the long-time fitted decay rates of the autocorrelations in Fig.~\ref{Hid}(b).
The results do show exponentially slow heating on top of a constant background decay, due to experimental imperfections such as decoherence, pulse imperfections, and higher-order terms in engineering the dipolar Hamiltonian $JD_y$.
In other words, these results demonstrate that while the Trotter error generically scales as $(J\tau)^2$~\cite{Heyl19,Lloyd96}, for quasiconserved quantities it only grows exponentially slow in $(J\tau)^{-1}$.
By normalizing the data to the data collected under the fastest drive ($J\tau=0.35$), the background decay is canceled, and the resulting dynamics only arises from the coherent evolution, as shown in Fig.~\ref{Hid}(c). Here we show the autocorrelation after the prethermal transient dynamics ($n\geq16$), as a function of the driving rate, $J\tau$.
In the absence of Floquet heating, the curves in Fig.~\ref{Hid}(c) would not change with $n$. Instead, the curves drop slowly when increasing $n$, qualitatively indicating that the system is still absorbing energy from the driving, and evolves toward the fully thermalized state at infinite temperature~\cite{Yin2020}.
At fixed $n$, the autocorrelation is close to 1 for small $J\tau$, but it decays to zero for larger $J\tau$ due to the presence of higher orders in $H_{pre}$ and the ultimate breakdown of the prethermal Hamiltonian picture.

\begin{figure*}[!htp]
\includegraphics[width=1.0\textwidth]{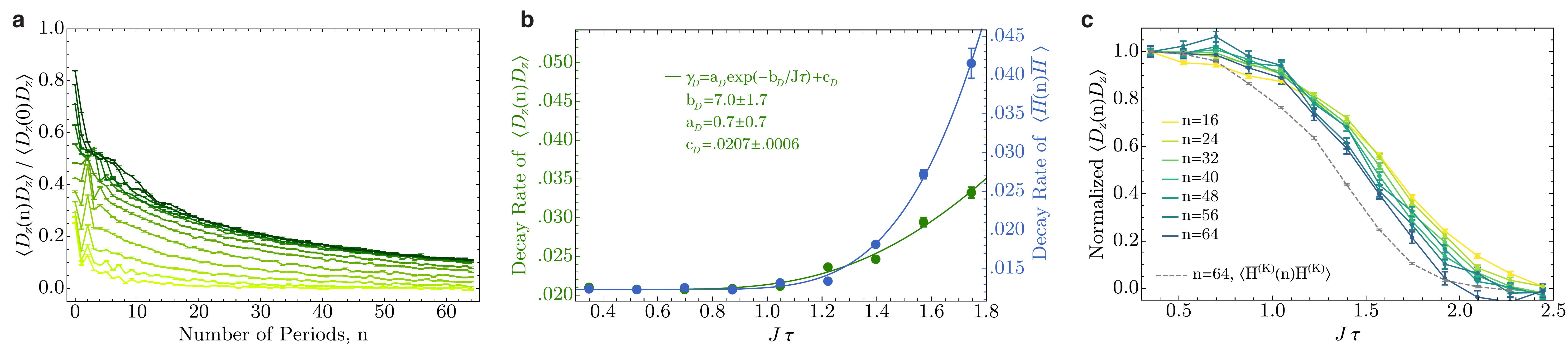}
\caption{\label{DzDz_KDM}
\textbf{Robustness of dipolar order.}
(a) Autocorrelation of $D_z$ in the kicked dipolar model with $J\tau$ from 0.35 (darker color) to 2.27 (lighter color) in steps of 0.175.
(b) We fit the autocorrelations in (a) from $n=20$ to $n=64$ to an exponentially decaying function $\exp(-\gamma_D n)$ and plot the decay rate $\gamma_D$ (left axis, green). The length of the error bars corresponds to two standard deviation of the fitted decay rate. The solid curve indicates the fit to $\gamma_D=a_D\exp(-b_D/J\tau)+c_D$. The fitted coefficients $a,b,c$ are shown in the plot with the 95\% confidence interval. The blue curves (right axis) are from Fig.~\ref{Hid}(b). The left and right axis have the same scale for easier comparison.
Comparing the fitted coefficient $b$ for both quasiconserved quantities reveals that $D_z$ has a slower dependence on $J\tau$.
% (c) Floquet heating rate of $\langle \overline H(n)\overline H\rangle$ (blue) and $\langle D_z(n) D_z\rangle$ (green). Dots are measured data and solid curves are fitted results as in Fig.~\ref{Hid}(b) and Fig.~\ref{DzDz_KDM}(b).
(c) Autocorrelation of $D_z$ versus $J\tau$ for different periods $n$. Lighter colors represent smaller $n$ and darker colors represent larger $n$. For a given $n$, the autocorrelation is normalized by $\langle D_z(n)D_z\rangle$ at $J\tau=0.35$, i.e. the leftmost point is normalized to 1. The grey dashed curve shows $\langle \overline H(n)\overline H\rangle$ at $n=64$.
% (e) Comparison of normalized $\langle D_z(n)D_z\rangle$ (gree) and $\langle \overline H(n)\overline H\rangle$ (blue) both at $n=64$.
In (a) and (c), error bars are determined from the noise in the NMR free induction decay~\cite{SOM}.
}
\end{figure*}

It is interesting to investigate whether this behavior is limited to energy quasiconservation, or if it occurs for other, non-trivial quasiconserved observables. Intriguingly, we find that such quasiconserved quantities not only exist, but can be even more robust than the prethermal energy.
In the kicked dipolar model, there is an additional quasiconserved quantity $D_{pre}\approx D_z$, which we call dressed dipolar order~\cite{Jeener67}.
We experimentally measured the autocorrelation of $D_z$ [Fig.~\ref{DzDz_KDM}(a)],
whose decay rate can be fitted to a constant background decay plus a term exponentially slow in $J\tau$ [Fig.~\ref{DzDz_KDM}(b)].
Surprisingly, we find that not only does $D_z$ have a smaller overall decay rate than $\overline H$, but the decay rate even shows a slower scaling with $J\tau$.
As a result, after normalizing by the background decay, the autocorrelation of $D_z$ is larger than that of $\overline{H}$ [Fig.~\ref{DzDz_KDM}(c)], indicating that for relatively large $J\tau$, $D_z$ is conserved for longer times than $\overline H$~\cite{SOM}, as also confirmed numerically in Extended Data.
In other words, there is a regime where the stroboscopic evolution can no longer be described by a static prethermal Hamiltonian, but still exhibits emergent symmetries, here the dressed dipolar order.
Although Ref.~\cite{Else17,Abanin17b,Wei19} have shown that $D_z$ is conserved by the static average Hamiltonian $\overline{H}$, this cannot explain its extraordinary robustness here, as indeed even for relatively large $\tau$ we can derive (see methods) the quasi-conservation law without first transforming to a static Hamiltonian. This could further lead to Floquet phases that have no static counterpart.

% \section{Conclusion}\label{sec:conclusion}
In conclusion, we studied Floquet prethermalization and heating in an interacting many-body system provided by a solid-state NMR quantum simulator, introducing a control protocol that can isolate Floquet effects from other experimental imperfections and decoherence.
Periodic driving is a powerful tool for quantum simulation and to induce novel phases of matter due to the Floquet dynamics. Whether such phases and engineered Hamiltonians can survive for long-enough times to allow interesting quantum simulations is a critical question for practical applications.
% Conversely, the final fate of the system, whether it thermalizes to infinite temperature, is a central issue in quantum thermodynamics.
Here we first observed the dynamics of a non-equilibrium state and showed that it indeed relaxes to a long-time steady-state given by the canonical ensemble of the prethermal Hamiltonian.
By measuring the dynamics of prethermal quasiconserved quantities, while keeping fixed experimental imperfections, we further revealed that the system final thermalization to infinite temperature happens with an exponentially small heating rate.
We succeeded in measuring to leading order not only the autocorrelation of the quasiconserved prethermal Hamiltonian $H_{pre}$, but also another emergent quasiconserved quantity, the dressed dipolar order $D_{pre}$.
Surprisingly, we find that the heating rate of $D_{pre}$ is smaller than for $H_{pre}$, with both rates scaling exponentially in the driving frequency. This result suggests that Floquet systems may exhibit conservation laws even when the dynamics can no longer be described by a static Hamiltonian.
Our work not only provides experimental evidence of Floquet prethermalization theory, but also opens new avenues for robust Floquet engineering and long-lived Floquet phases of matter.

\textit{Note added.} During the preparation of this
manuscript, we became aware of related experiments about Floquet prethermalization in Bose-Hubbard model~\cite{Antonio2020}.

\textbf{Acknowledgments}
The authors would like to thank H. Zhou, W.-J Zhang and Z. Li for discussion. This work was supported in part by the National Science Foundation under Grants No. PHY1734011, No. PHY1915218, and No. OIA-1921199.

\textbf{Author contributions}
P.P. performed the experiments. P.P. and C.Y. analyzed the data and developed the theoretical modeling. P.C. supervised the project. All authors discussed the results and contributed to the manuscript.

\textbf{Competing interests}
The authors declare no competing interests.

%\end{acknowledgments}

\bibliography{Biblio}

\clearpage
\section*{Methods}
\subsection*{Experimental System}
We use nuclear spins in fluorapatite as our experimental testbed.
{The $^{19}$F spins-1/2 form linear chains [Fig.~\ref{ZZ_ZDy}(b)] in the crystal and interact with each other via magnetic dipolar interactions. A single crystal is placed in a 7~T magnetic field at room temperature.
In such a strong magnetic field the interaction Hamiltonian for the $^{19}$F spins is given by the secular dipolar Hamiltonian $H_\text{Dipz}=J_0 D_z$,
%+\sum_j h_j S_z^j$,
where $D_z=\sum_{j<k} \frac{1}{2}\left(3S_z^j S_z^k - \vec{S}_j\cdot\vec{S}_k \right)/r_{jk}^{3},
$ with $r_{jk}=|k-j|$ the normalized distance between two spins and $J_0=-29.7$~krad/s the nearest neighbor coupling strength. Here \(S_\alpha^j\) \((\alpha=x,y,z)\) are spin-1/2 operators of the \(j\)-th $^{19}$F spin and $\vec{S}_j=(S_x^j,S_y^j,S_z^j)^T$
(see~\cite{SOM} for more details).
In the short timescale, the system can be approximated as an ensemble of identical spin chains~\cite{ Ramanathan11}, because the interchain coupling is $\sim\!40$ times weaker than the intrachain coupling.
The quasi 1D nature of the crystal allows us to compare with numerical simulation easily, but the results of this paper should be generic to any dimension with short-range interactions.
The NMR signal is summed over a macroscopic number of chains in the crystal. The average chain length $L$ is larger than 50, much longer than the extent of correlations in the experiments.
This large system size enables studying thermalization, a process that only happens in the thermodynamic limit.
On the timescales explored the $^{19}$F spin system can be considered a closed system, as the coupling to $^{31}$P spins in the lattice is refocused by the applied control, and the spin-lattice relaxation effects are negligible ($T_1\approx0.8$~s).
Such a long relaxation time allows us to resolve the exponentially slow heating rate.
{We can thus model the $^{19}$F spins by the closed quantum dynamics of spins interacting via dipolar couplings.
}}

{The initial state of a room-temperature NMR experiment with $L$ spins is described by the density matrix $\rho_0\!\approx\!(\mathbb{1}\!-\!\epsilon Z)/2^L$, with $\epsilon=\beta\omega_0\! \sim\! 10^{-5}$, where $\omega_0$ is the Zeeman energy and $\beta$ the inverse temperature.
%and $Z=\sum_j S_z^j$ the collective magnetization operator along z-axis.
As the identity operator describes a totally mixed state that does not produce any NMR signal, we only care about the deviation $\delta\rho_0=Z$. The NMR spectrometer measures the collective transverse magnetization, but with collective control we can measure the magnetization around any axis. In addition, the Jeener-Broekaert pulse pair~\cite{Jeener67} can be used to evolve the collective magnetization into the dipolar ordered state, $D_z$, plus some highly correlated operators which do not contribute to the signal.
Then, both the initial state $\delta\rho_0$ and the observable $\mathcal{O}$ can be chosen to be the collective magnetization operator $\sum_jS^j_\alpha$ or dipolar order operator $D_\alpha=\sum_{j<k} \frac{1}{2}\left(3S_\alpha^j S_\alpha^k - \vec{S}_j\cdot\vec{S}_k \right)/r_{jk}^3$ with $\alpha$ being an arbitrary direction.
The signal we measure can be rewritten as the two-point correlation at infinite temperature, $\text{Tr}[U(t)\delta\rho_0 U^{\dag}(t)\mathcal{O}]/2^{L}\equiv \langle \delta\rho(t)\mathcal{O}\rangle_{\beta=0}$.
That is, we are effectively measuring the correlation of a system at infinite temperature where (the deviation of) the density matrix becomes the time-dependent observable.
% That is, in our experiments, (the deviation of) the density matrix plays the role of an observable for an effective simulated system at infinite temperature.
Here and in the main text, we drop the subscript $\beta=0$ for brevity.}

\subsection*{Prethermal expansion}
\label{app:Dz}
\subsubsection{Fast driving expansion}
 For any periodic Hamiltonian $H(t)\!=\!H(t+\!\tau)$, the stroboscopic propagator $U_F=\mathcal{T}\left[ \exp\left(-i\int_{0}^\tau H(t)dt\right)\right]$ (with $\mathcal{T}$ the time-ordering operator) can be written in terms of a time-independent \textit{Floquet Hamiltonian} $H_F$, $U_F=\exp(-iH_F\tau)$. $H_F$ can be expanded in powers of $\tau$ by the Floquet-Magnus expansion~\cite{Magnus54, Blanes09}.
 In interacting many-body systems, the expansion typically diverges and a quasilocal $H_F$ cannot be found~\cite{Bukov15, Blanes09}, indicating there is no energy conservation and the system eventually heats up to infinite temperature. Still, in some cases a prethermal Hamiltonian
$H_{pre}$ emerges from the truncated Floquet-Magnus expansion
\begin{equation}\label{eq:Hpre}
H_{pre}=\sum_{m=0}^{m^*}\tau^m \Omega_m.
\end{equation}
Here the zeroth order term is the average Hamiltonian $\Omega_0\!=\!\overline H\!=\!1/\tau\int_{0}^\tau\! H(t)dt$ and higher order term $\Omega_m$ involves $m$ nested commutators, e.g., $\Omega_1\!=\!-(i/2\tau)\int_{0}^\tau dt_1 \int_{0}^{t_1} dt_2 \left[H(t_1),H(t_2)\right]$.
For a prethermal Hamiltonian, the residual time-dependent part $\delta H(t)=H_F-H_{pre}$ is exponentially small, $\|\delta H(t)\|<\exp\left(-O(1/\tau)\right)$ (where $||\cdot||$ denotes norm of local terms).
 Then, the system dynamics is governed by this truncated Floquet Hamiltonian up to an exponentially long time $t^*\sim\exp(O(1/\tau))$, before finally reaching infinite temperature.

\subsubsection{Large kicking field expansion}
We can use this prethermal formalism~\cite{Else17,Kuwahara16} to theoretically demonstrate that the kicked dipolar model leads to prethermal quasiconserved quantities featuring exponentially slow heating.
With a local unitary transformation $e^{-S}$, we can rewrite the Floquet operator for the kicked dipolar model as
\begin{equation}\label{eq:pre0}
e^S \lr{ e^{-i hZ\tau}e^{-iJD_y\tau} } e^{-S} = e^{-i h\widetilde{Z}\tau} e^{-i\tau (\widetilde{D}+\delta \widetilde{H})},
\end{equation}
where $[\widetilde{Z},\widetilde{D}]\!=\!0$ and $\delta \widetilde{H}$ is exponentially small in $\min\{O(h/J),O(1/h\tau)\}$ (we use $\widetilde{\mathcal{O}}$ to denote operators in the new frame.) $\widetilde{D}$ is then exponentially conserved, yielding a prethermal quasiconserved quantity $D_{pre}=e^{-S}\widetilde{D}e^{S}$ in the original frame.

To find $S$, we first rewrite the transformation in Eq.~(\ref{eq:pre0}) as
\begin{equation}\label{eq:pre}
e^{i \epsilon hZ\tau} e^S e^{-i \epsilon hZ\tau} e^{-i \epsilon^2 JD_y\tau} e^{-S} = e^{-i\tau (\widetilde{D}+\delta \widetilde{H})},
\end{equation}
where we drop the tilde on $Z$ because the two operators have the same matrix representation.
Here $\epsilon=1$ is the case at hand, but we will evaluate Eq.~(\ref{eq:pre}) as a perturbation in the small parameter $\epsilon\ll1$. In particular, we set both $h\tau$ and $J/h$ to be small numbers of order $\epsilon$.
After expanding the operators, $\widetilde{D}=\epsilon D_1+\epsilon^2 D_2+\cdots$, $S = \epsilon S_1+\epsilon^2 S_2+\cdots$, one can collect terms that are of order $\epsilon^{j}$ on both sides of Eq.~(\ref{eq:pre}), using the Magnus expansion to evaluate the products of exponentials.
The $j^{th}$ order is given by
\begin{equation}\label{eq:pre_j}
 -i\tau D_j = \left[S_{j-1},-ihZ\tau\right] + h_j,
\end{equation}
where $h_j$ only contains $hZ\tau$, $JD_y\tau$ and $S_{j'<j}$. Higher orders can be found recursively from $h_1\!=\!0$ and $h_2\!=\!-iJD_y\tau$. %(e.g, $h_3=([S_1,[S_1,-ihZ\tau]]+[ihZ\tau,[ihZ\tau,S_1]])/2+[S_1,-iJD_y\tau]$).
Assuming all orders $S_{j'}$ with $j'<j-1$ are known, we determine $S_{j-1}$ by requiring $\left[S_{j-1},-ihZ\tau\right]$ to cancel the terms in $h_j$ that do not commute with $Z$. This can be conveniently obtained by
decomposing $h_j = \sum_{q=0,\pm1,\cdots} h_{jq}$ such that $[Z, h_{jq}] = q h_{jq}$ ($h_{jq}$ is called the $q^\text{th}$ quantum coherence of $Z$~\cite{Wei18,Munowitz75,Garttner18}.) Eq.~(\ref{eq:pre_j}) is satisfied by setting $-i \tau D_j = h_{j0}$ and $S_{j-1} = i\sum_{q\neq 0} h_{jq}/(hq\tau)$. This procedure results in a localized expansion:
 for nearest-neighbor interaction, the range of $S_j$ is at most $j$, yielding an exponentially localized $e^S$ (similar localization is expected for short-range interactions found in our experiments.)
Truncating the expansion so it remains convergent and local
leads to the exponentially small residual $\delta \widetilde{H}$.

In the $\tau\to0$ limit, the $S_j$ operators are dominated by the $(J/h)^{j}$ term, and the Floquet quasiconserved quantities agree with the prethermal quasiconserved quantities of the static Hamiltonian $\overline{H}$~\cite{Wei19,Else17}. In this regime, as in the static case, the expansion is a series in $J/h$, converging for $h/J \gtrsim 0.5$ with an error $\delta \widetilde{H}\approx \exp[-O(h/J)]$, which yields $D_{pre}=- \frac{1}{2}D_z +O((J/h)^2)$~\cite{Wei19}.
%(Note that here in experiments and simulations we used $h/J=1$).

Instead, for relatively larger $h\tau$, the $S_j$ operators are dominated by $(h\tau)^j$ and $\delta \widetilde{H}\approx \exp(-O(1/h\tau))$, in agreement with the exponentially slow Floquet heating. This prethermal expansion is a generalization of earlier results~\cite{Else17}, which required $[\widetilde{D},e^{-ih\widetilde{Z}\tau}]=0$ for $h\tau=\pi$ to realize discrete time crystals. Here instead we impose the stronger requirement $[\widetilde{D},\widetilde{Z}]=0$, and the expansion is valid for arbitrary $h\tau$. Note that since the right-hand side of Eq.~\ref{eq:pre0} still describes a Floquet system, the quasiconservation is derived without first transforming to a static Hamiltonian.

Numerical simulations~\cite{Yin2020} of the series convergence and the infinite-time correlation (see extended data) suggest that $D_{pre}$ is more robust than $H_{pre}$, in agreement with the experimental results in the main text.

\setcounter{figure}{0}
\renewcommand{\figurename}{\textbf{Extended Data Fig.}}

\begin{figure*}
\centering
\includegraphics[width=1.0\textwidth]{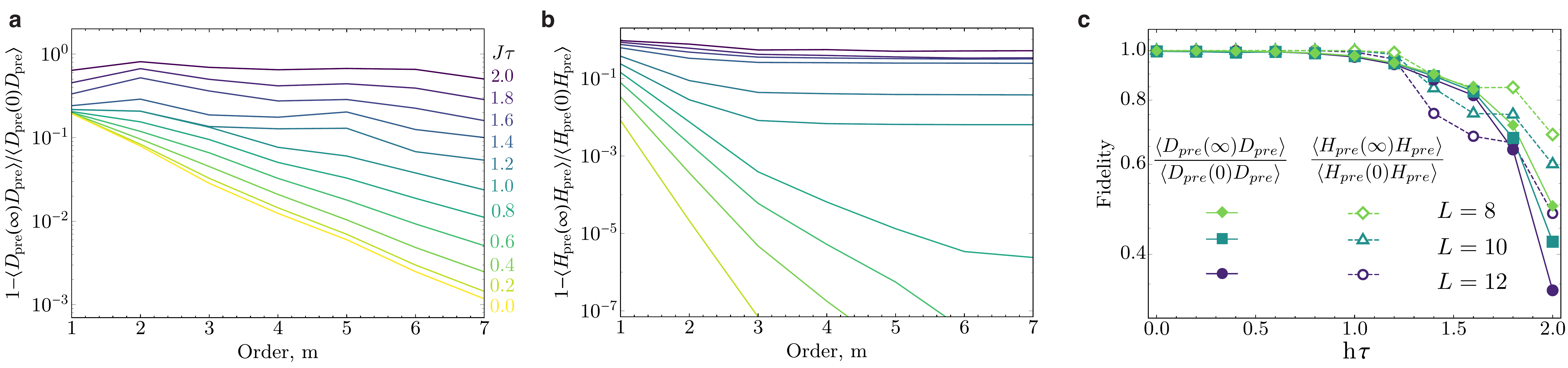}
\caption{\label{Dpre_theory}
\textbf{Signatures of a robust quasiconserved quantity for the kicked dipolar model.} (a) We numerically evaluate the expansion of of the quasiconserved quantity $D_{pre}$ and plot the
norm of each term (normalized by $L2^L$) as a function of the order $m$, for various $h\tau=J\tau$. $J\tau$ varies from 0 (light colors) to 2 (dark colors) in steps of 0.2.
(a) Infidelity $1-\langle D_{pre}(\infty)D_{pre} \rangle/\langle D_{pre}D_{pre} \rangle$ of the infinite-time averaged $D_{pre}$ as a function of the order, $m$, for various $h\tau=J\tau$. $J\tau$ varies from 0 (light colors) to 2 (dark colors) in steps of 0.2.
(b) Infidelity $1-\langle H_{pre}(\infty)H_{pre} \rangle/\langle H_{pre}H_{pre} \rangle$ of the infinite-time averaged prethermal Hamiltonian $H_{pre}$ as a function of the order, $m$, for various $h\tau=J\tau$.
$L=12$ was used in (a-b).
The normalized autocorrelation of $H_{pre}$ converges to 1 in a smaller parameter range ($J\tau\lesssim 1$) than $D_{pre}$ ($J\tau\lesssim 1.6$)
(c) Fidelities of the two conserved quantities ($\langle H_{pre}(\infty)H_{pre} \rangle/\langle H_{pre}H_{pre} \rangle$ and $\langle D_{pre}(\infty)D_{pre} \rangle/\langle D_{pre}D_{pre} \rangle$) evaluated to $7^\text{th}$ order as a function of $h\tau$ for three different system sizes. The fidelities show a significant drop when $L$ is increased from 8 to 12 at $J\tau\gtrsim1.8$ for $D_{pre}$ and $J\tau\gtrsim1.2$ for $H_{pre}$, again indicating that $D_{pre}$ is more robust than $H_{pre}$.
}
\end{figure*}

%\bibliography{Biblio}

\clearpage
\newpage

\clearpage
\begin{center}
\Large\textbf{Supplemental}\normalsize	
\end{center}

\setcounter{figure}{0}
\renewcommand{\figurename}{\textbf{Fig.}}

\section{Experimental background decay rate as a function of $J\tau$} \label{app:background}
In the main text we measured the Floquet heating for a periodic, Hamiltonian switching scheme. While it would be easy to change the period by increasing the time between switches, this would lead to experiments performed with different total times or a different number of control operations. In turns, this can introduce variable amount of decoherence and relaxation effects, and of control errors. Instead, we kept the time for one Floquet period constant and used Hamiltonian engineering to vary the Hamiltonian strength in order to vary the Floquet driving frequency.

\begin{figure}[t]
\centering
\includegraphics[width=0.98\columnwidth]{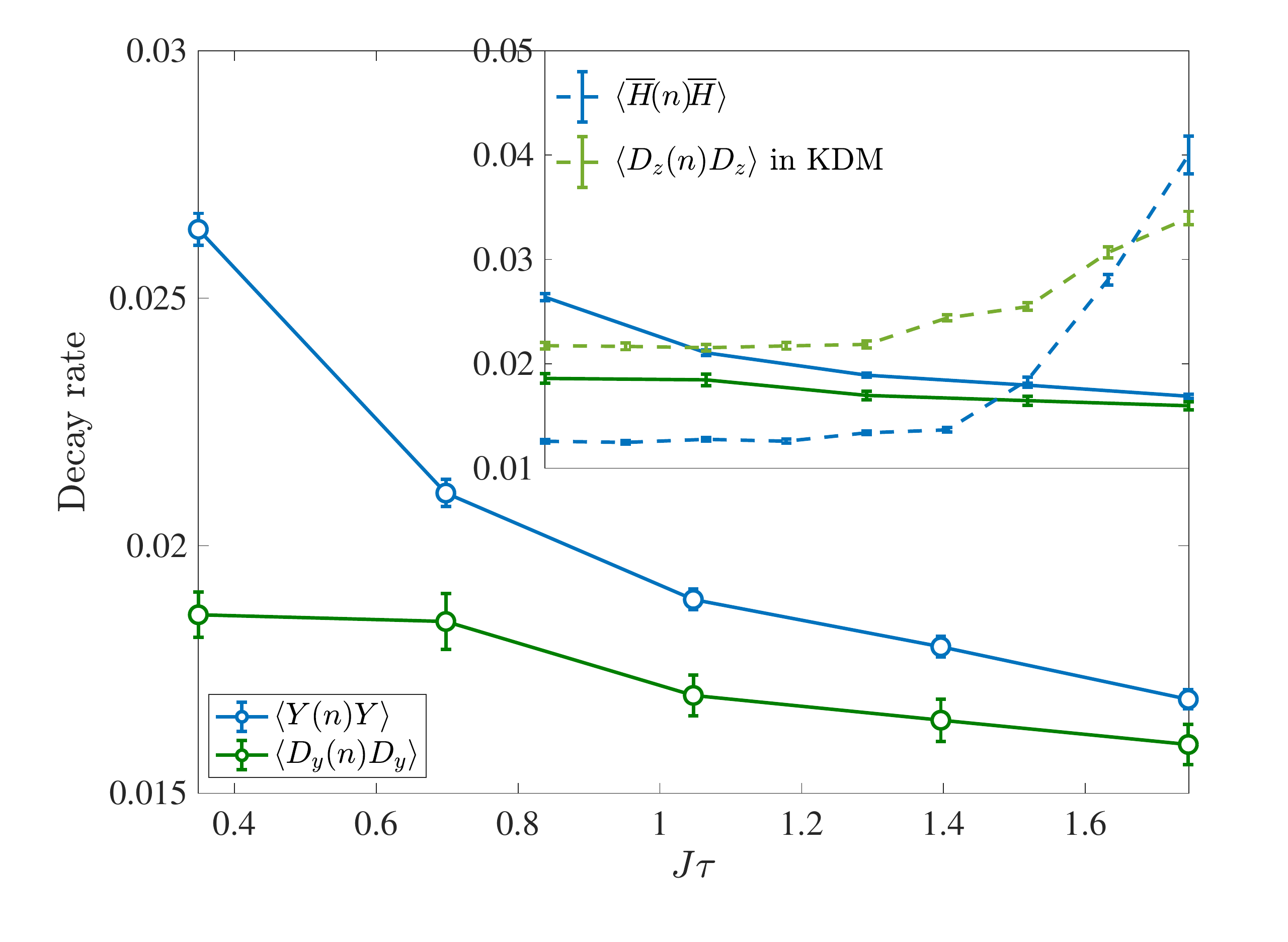}
\caption{\label{fig:bgDR}
Decay rate of $\langle Y(n)Y\rangle$ (blue) and $\langle D_y(n)D_y\rangle$ (green) under engineered dipolar Hamiltonian $JD_y$ as a function of $J\tau$. {The range of $J\tau$ studied was obtained by varying the scaling $u$ (see Sec.~\ref{app:Ham}) from 0.098 to 0.646, while keeping fixed $\tau=120\mu s$. In the inset, we compare the background decay rates with the Floquet decay rates (dashed lines).}
}
\end{figure}

One of the assumptions in our work is that the background decay rate does not change much with driving frequency (compared to the change in Floquet heating rate). In this section, we provide experimental evidence for this assertion.
When changing driving frequency, we are changing (i) the effective strength $J$ of the engineered dipolar interaction $JD_y$ and (ii) the kicking angle in the kicked dipolar model by a phase shift (see \ref{app:Ham}).
As phase shift angles are usually very accurately implemented in NMR experiments, we focus on the engineered dipolar interaction, which is obtained by Floquet engineering itself, as explained in \ref{app:Ham}.
To quantify how good is the engineered $JD_y$, we measure $\langle Y(n)Y\rangle$ and $\langle D_y(n)D_y\rangle$ under the engineered Hamiltonian $JD_y$, without kicking field nor direction alternation, as shown in Fig.~\ref{fig:bgDR}.
Qualitatively, we observe that the decay rate \textit{decreases} when increasing $J\tau$, while the decay rate when performing Hamiltonian switching [Fig.~\ref{Hid}(b) and Fig.~\ref{DzDz_KDM}(b) of the main text] increases with $J\tau$.
This opposite trend even strengthens our conclusions in the main text: (i) the decay of autocorrelations is indeed due to Floquet heating, but not to experimental errors, (ii) the decay rate curve shown in Fig.~\ref{Hid}(b) and Fig.~\ref{DzDz_KDM}(b) should be steeper, thus further isolating the exponential scaling from other possibilities (e.g. quadratic prethermalization due to first order perturbation), (iii) the change of Floquet heating rate is more evident in magnetization than in dipolar state, thus our conclusion that the $D_z$ quasiconservation lives longer than $\overline H= hZ+JD_y$ is even stronger.

Note that the maximum difference between the decay rate of $\langle D_y(n)D_y\rangle$ over the range of $J\tau$ considered is $\sim 0.003$, much smaller than the Floquet heating rate in the main text. A quantitative analysis is challenging because the specific form of error terms is unknown, and $JD_y$ is an interacting Hamiltonian thus error accumulation is intractable.
Here we use some simple arguments to argue that variations in the background decay with $J\tau$ have little to no influence on our results. First, we note that while in the main text we are interested in the decay of the autocorrelation of $H_{pre}$ and $D_{pre}$, here with $H=JDy$ we can only discuss the decay of $D_y$ and $Y$, since other quantities that are not conserved display very fast decay which is not informative.
For example, in the main text we measure $D_z$, which thermalizes even under the ideal $D_y$ and thus we cannot distinguish thermalization from decay due to experimental imperfections in the engineered dipolar Hamiltonian $D_y$. Still, as $D_z$ and $D_y$ overlap, if the background decay of $D_z$ had a significant change with $J\tau$, it would be reflected in $D_y$, which is not observed. Therefore, we expect the change in the background decay rate for $\langle D_z(n) D_z\rangle$ to be small as well.
Here we can only probe the background decay rate of $Y$, while in the main text we are interested in the longitudinal magnetization, $Z$, that appears in $\langle \overline H(n)\overline H\rangle$ [see Fig.~\ref{Hid}(b)]. The transverse magnetization decay rate is, however, a upper bound for $Z$, since in NMR experiments $Z$ is usually more robust against errors than $Y$ due to the large magnetic field in z-axis that suppresses decoherence and experimental errors that do not conserve the total Zeeman energy (we note that we typically do not explicitly write the Zeeman energy in the Hamiltonians as we work in the rotating frame).
Even if the variation in the background decay for $Z$ were as large as what observed for $Y$ in these experiments ($\sim0.009$), it would still be still small compared with Floquet (see inset of Fig.~\ref{fig:bgDR}). In addition, in the kicked dipolar model, we can consider $ D_y$ as being subjected to rotations along $Z$ that further cancel out the error terms in the engineered $JD_y$ that do not conserve $Z$.
As a result, the decay rate of $Y$ due to the engineered $D_y$ is larger, by about a factor of 2, than the baseline decay of $\langle \overline H(n)\overline H\rangle$ in the kicked dipolar model (they are 0.254 and 0.123, respectively, in the fastest driving case $J\tau=0.35$).

\section{Experimental System, Control and Data Analysis}
\subsection{Experimental System}
\label{app:exp}
%todo:SM (expt)
The system used in the experiment was a single crystal of fluorapatite (FAp). Fluorapatite is a hexagonal mineral with space group \(P6_3/m\), with the \(^{19}\)F spin-1/2 nuclei forming linear chains along the \(c\)-axis. Each fluorine spin in the chain is surrounded by three \(^{31}\)P spin-1/2 nuclei.
We used a natural crystal, from which we cut a sample of approximate dimensions 3 mm$\times$3 mm$\times$2 mm.
The sample is placed at room temperature inside an NMR superconducting magnet producing a uniform $B=7$ T field. The total Hamiltonian of the system is given by
\begin{equation}
H_\mathrm{tot}=\omega_F \sum_k S_z^k+\omega_P \sum_\kappa s_z^\kappa+H_{F}+H_P+H_{FP}
\label{eq:Hamtot}	
\end{equation}
The first two terms represent the Zeeman interactions of the F($S$) and P($s$) spins, respectively, with frequencies $\omega_F=\gamma_FB\approx (2\pi)282.37$ MHz and $\omega_P=\gamma_PB=(2\pi)121.51$ MHz, where $\gamma_{F/P}$ are the gyromagnetic ratios. The other three terms represent the natural magnetic dipole-dipole interaction among the spins, given generally by
\begin{equation}
 H_\mathrm{dip}=\sum_{j<k}\frac{\hbar\gamma_j\gamma_k}{|\vec r_{jk}|^3}\left[\vec S_j\cdot\vec S_k-\frac{3\vec S_j\cdot\vec r_{jk}\,\vec S_k\cdot\vec r_{jk}}{|\vec r_{jk}|^2}\right],
\end{equation}
where $\vec r_{ij}$ is the vector between the $ij$ spin pair. Because of the much larger Zeeman interaction, we can truncate the dipolar Hamiltonian to its energy-conserving part (secular Hamiltonian). We then obtain the homonuclear Hamiltonians
\begin{equation}
 \begin{aligned}
 H_F&=\frac{1}{2}\sum_{j<k}J^F_{jk}(2 S_z^j S_z^{k}- S_x^j S_x^{k}- S_y^j S_y^{k}) \\ H_P&=\frac{1}{2}\sum_{\lambda<\kappa}J^P_{\kappa\lambda}(2s_z^\lambda s_z^{\kappa}-s_x^\lambda s_x^{\kappa}-s_y^\lambda s_y^{\kappa})
 \end{aligned}
\end{equation}
and the heteronuclear interaction between the $F$ and $P$ spins,
\begin{equation}
 H_{FP}=\sum_{k,\kappa} J^{FP}_{k,\kappa}S_z^ks_z^\kappa,
\end{equation}
with $J_{jk}=\hbar\gamma_j\gamma_k\frac{1-3\cos(\theta_{jk})^2}{|\vec r_{jk}|^3}$, where $\theta_{jk}$ is the angle between the vector $\vec r_{jk}$ and the magnetic field $z$-axis. The maximum values of the couplings (for the closest spins) are given respectively by $J^F=-32.76$ krad s$^{-1}$, $J^P=1.20$ krad s$^{-1}$ and $J^{FP}=6.12$ krad s$^{-1}$.
\begin{figure}[b]
\centering
\includegraphics [width=0.98\linewidth ]{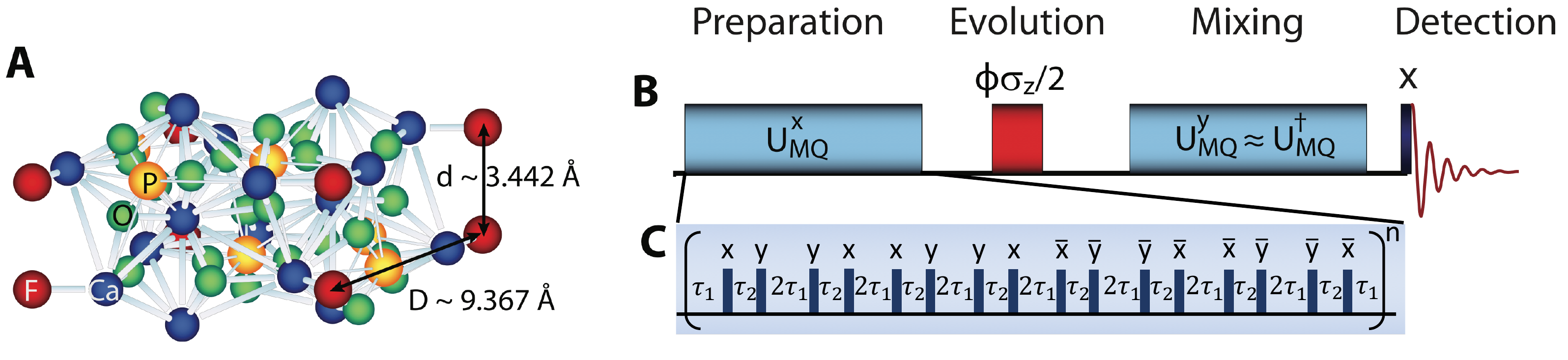}
\caption{\textbf{A} Fluorapatite crystal structure, showing the Fluorine and Phosphorus spins in the unit cell. \textbf{B} NMR scheme for the generation and detection of MQC. In the inset (\textbf{C}) an exemplary pulse sequence for the generation of the $H_\mathrm{dipy}$. Note that thanks to the ability of inverting the sign of the Hamiltonian, the scheme amounts to measuring out-of-time order correlations.
}\label{fig:mqcd}
\end {figure}

The dynamics of this complex many-body system can be mapped to a much simpler, quasi-1D system. First, we note that when the crystal is oriented with its $c$-axis parallel to the external magnetic field
the coupling of fluorine spins to the closest off-chain fluorine spin is $\approx40$ times weaker, while in-chain, next-nearest neighbor couplings are $8$ times weaker.
 Previous studies on these crystals have indeed observed dynamics consistent with spin chain models, and the system has been proposed as solid-state realizations of quantum wires
~\cite{Cappellaro07l,Cappellaro11,Ramanathan11}. This approximation of the experimental system to a 1D, short-range system, although not perfect has been shown to reliably describe experiments for relevant time-scales~\cite{RufeilFiori09b,Zhang09}. The approximation breaks down at longer times, with a convergence of various effects: long-range in-chain and cross-chain couplings, as well as pulse errors in the sequences used for Hamiltonian engineering. In addition, the system also undergoes spin relaxation, although on a much longer time-scale ($T_1=0.8~$s for our sample).

\subsection{Error analysis}\label{app:err}
\begin{figure}[h]
\centering
\includegraphics[width=90mm,clip]{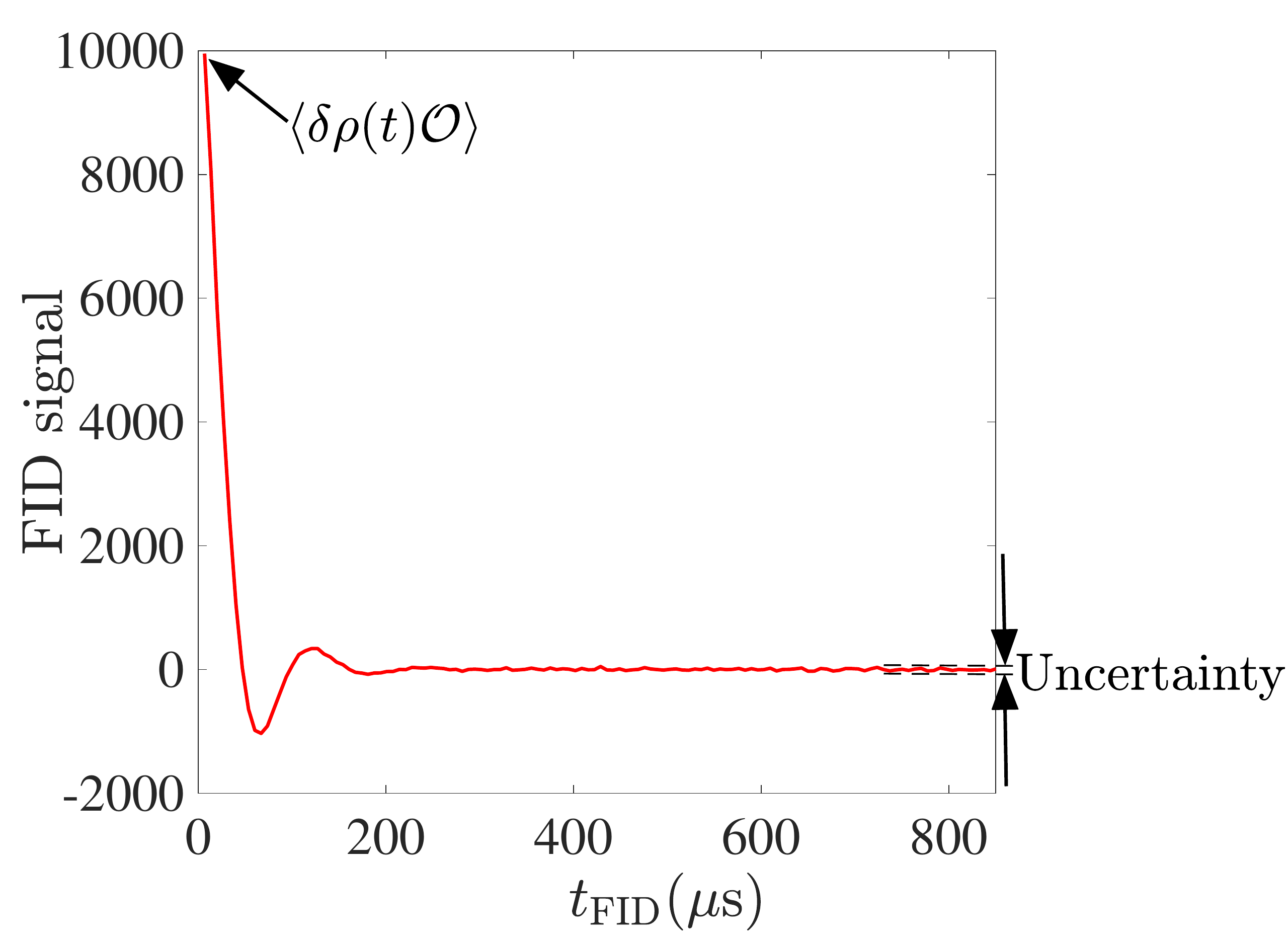}
\caption{\label{fig:FID}
An example of FID. 128 data points are taken in total. The first data point gives $\langle(\delta\rho(t)\mathcal{O}\rangle$ and the standard deviation of the last 20 points gives the uncertainty of $\langle(\delta\rho(t)\mathcal{O}\rangle$.
}
\end{figure}

%todo:SM (error)
In experiments, we want to measure the correlation $\langle\delta\rho(t)\mathcal{O}\rangle$, where $\delta\rho(t)= U(t)\delta\rho(0)U(t)$ is the nontrivial part of the density matrix evolved under a pulse-control sequence for a time $t$.
Instead of just performing a single measurement after the sequence,
 %To get the uncertainty of $\langle(\delta\rho(t)\mathcal{O}\rangle$, instead of just measuring one point,
we continuously monitor the free evolution of $\delta\rho(t)$ under the natural Hamiltonian $H_\mathrm{dip}$, from $t$ to $t+t_{\textrm{FID}}$.
The measured signal is called in NMR free induction decay (FID) and a typical FID trace is shown in Fig. \ref{fig:FID}).
This signal trace allows us to extract not only the amplitude of the correlation (from the first data point) but also its uncertainty.
We take the standard deviation of the last 20 data points in the FID as the uncertainty of the $\langle\delta\rho(t)\mathcal{O}\rangle$. This uncertainty is used with linear error propagation to obtain the error bars of all the quantities analyzed in the main text.

\subsection{Hamiltonian Engineering}\label{app:Ham}
%todo: SM(Ham eng.)
In the main text we focused on the Floquet heating (Trotter error) for a periodic alternating scheme, switching between two Hamiltonians. In order to avoid longer times and/or different numbers of control operations when changing the Trotter step (Floquet period), we engineered Hamiltonians of variable strengths. Then, the Hamiltonians themselves are obtained stroboscopically by applying periodic rf pulse trains to the natural dipolar Hamiltonian that describes the system, and are thus themselves Floquet Hamiltonians. Since we only varied the sequences, but not the Floquet period, this step does not contribute to the behavior described in the main text, as we further investigate in Sec.~\ref{app:background}.

We used Average Hamiltonian Theory (AHT~\cite{Haeberlen68}) as the basis for our Hamiltonian engineering method, to design the control sequences and determine the approximation errors.
The dynamics is induced by the total Hamiltonian \(H=H_\text{dip}+H_\text{rf}\),
where \(H_\text{dip}=\frac{1}{2}\sum_{j<k}J_{jk}(2 S_z^j S_z^{k}-S_x^j S_x^{k}-S_y^j S_y^{k})+\sum_j h_j S_z^j\) is the system Hamiltonian,
and \(H_\text{rf}(t)\) is the external Hamiltonian due to the rf-pulses.
The density matrix \(\rho\) evolves under the total Hamiltonian according to \(\dot\rho=-i[H,\rho]\).
We study the dynamics into a convenient interaction frame, defined by \(\rho'={U_\text{rf}}^{\dagger}\rho U_\text{rf}\), where \(U_\text{rf}(t)=\mathcal{T}\exp[-i\int_0^t H_\text{rf}(t') dt']\) and \(\mathcal{T}\) is the time ordering operator.
In this \textit{toggling} frame, \(\rho'\) evolves according to \(\dot{\rho}'=-i[H(t),\rho']\), where \(H(t)={U_\text{rf}}^{\dagger}H_\text{dip} U_\text{rf}\).
Since \(U_\text{rf}\) is periodic, \(H(t)\) is also periodic with the same period $\tau$, and gives rise to the Floquet Hamiltonian, $H_F$, as as \(U(\tau)=\exp[-i H_F \tau]\).
Note that if the pulse sequence satisfies the condition \(U_\text{rf}(\tau)=1\), the dynamics of \(\rho\) and \(\rho'\) are identical when the system is viewed stroboscopically, i.e., at integer multiples of \(\tau\), where the toggling frame coincides with the (rotating) lab frame.
% The system evolves as if under a time-independent Hamiltonian \(H_\text{Flq}\).
% To calculate \(H_\text{Flq}\) we employ the Magnus expansion as is usual in AHT: \(H_\text{Flq}=H^{(0)}+H^{(1)}+\cdots\).
% The first two terms are given by
% \begin{align*}
% H^{(0)}=\frac{1}{t_c}\int_0^{t_c}H(t)dt ,\quad H^{(1)}=\frac{-i}{2t_c}\int_0^{t_c}dt_2\int_0^{t_2}dt_1[H(t_2),H(t_1)].
% \end{align*}
% The zeroth order of the average Hamiltonian \(H^{(0)}\) is often a good approximation to the Floquet Hamiltonian \(H_\text{Flq}\), as the first order can be set to zero by simple symmetrization of the pulse sequence.

We devised control sequences to engineer a scale-down, rotated version of the dipolar Hamiltonian~\cite{Wei18,Wei19}. We usually look for control sequences that would engineer the desired Hamiltonian up to second order in the Magnus-Floquet expansion.
Then, to engineer the interaction $D_y$, we use a 16-pulse sequence. The basic building block is given by a 4-pulse sequence~\cite{Kaur12,Yen83} originally developed to study MQC.
We denote a generic 4-pulse sequence as \(P(\tau_1,{\bf n}_1,\tau_2,{\bf n}_2,\tau_3,{\bf n}_3,\tau_4,{\bf n}_4,\tau_5)\), where \({\bf n}_j\) represents the direction of the \(j\)-th \(\pi/2\) pulse, and \(\tau_j\)'s the delays interleaving the pulses. In our experiments, the \(\pi/2\) pulses have a width \(t_w\) of typically 1 \(\mu\)s. \(\tau_j\) starts and/or ends at the midpoints of the pulses (see also Fig.~\ref{fig:mqcd}). In this notation, our forward 16-pulse sequence can be expressed as
\begin{widetext}
\begin{gather*}
P(\tau_1,{\bf x},\tau_2,{\bf y},2\tau_1,{\bf y},\tau_2,{\bf x},\tau_1)P(\tau_1,{\bf x},\tau_2,{\bf y},2\tau_1,{\bf y},\tau_2,{\bf x},\tau_1)P(\tau_1,{\bf \overline{x}},\tau_2,{\bf \overline{y}},2\tau_1,{\bf \overline{y}},\tau_2,{\bf \overline{x}},\tau_1)P(\tau_1,{\bf \overline{x}},\tau_2,{\bf \overline{y}},2\tau_1,{\bf \overline{y}},\tau_2,{\bf \overline{x}},\tau_1)
\end{gather*}
and the backward sequence as
\begin{gather*}
P(\tau_3,{\bf y},\tau_3,{\bf x},2\tau_4,{\bf x},\tau_3,{\bf y},\tau_3)P(\tau_3,{\bf y},\tau_3,{\bf x},2\tau_4,{\bf x},\tau_3,{\bf y},\tau_3)P(\tau_3,{\bf \overline{y}},\tau_3,{\bf \overline{x}},2\tau_4,{\bf \overline{x}},\tau_3,{\bf \overline{y}},\tau_3)P(\tau_3,{\bf \overline{y}},\tau_3,{\bf \overline{x}},2\tau_4,{\bf \overline{x}},\tau_3,{\bf \overline{y}},\tau_3)
\end{gather*}
\end{widetext}
where \(\{{\bf \overline{x}},{\bf \overline{y}}\}\equiv \{{\bf -x},{\bf -y}\}\). The delays are given by
\begin{gather*}
\begin{aligned}
\tau_1&=\tau_0(1-u), \quad
\tau_2=\tau_0(1+2u), \\
\tau_3&=\tau_0(1+u), \quad
\tau_4=\tau_0(1-2u),
\end{aligned}
\end{gather*}
where \(\tau_0\) is 5 \(\mu\)s in this paper. The cycle time \(t_c\), defined as the total time of the sequence, is given by \(\tau=24\tau_0\). \(u\) is a dimensionless adjustable parameter, and is restricted such that none of the inter-pulse spacings becomes negative. To the zeroth order Magnus expansion, the above sequence realizes Hamiltonian $uJ_0D_y$ and $uJ_0=J$.

A uniform transverse field can be introduced in $H^{(0)}$
%in two ways. One strategy is to simply apply pulses off-resonance, so that the resulting \(H^{(0)}\) contains the term \(-g\Delta \omega /2\sum_j \sigma_z^j\), where \(\Delta \omega\) is the resonance offset. This approach is easy to implement, but it cannot achieve independent control over the disordered and uniform fields, and it can result in lower-quality pulses. We use a second approach which is based on
by phase-shifting the entire pulse sequence. Consider rotating the \(n\)-th cycle of the pulse sequence by \((n-1)\phi\) around the \({\bf z}\) axis, which can be accomplished by phase shifting all the pulse directions ${\bf n}_j$ in the \(n\)-th cycle by \((n-1)\phi\). The evolution operator for each cycle is given by
\begin{align*}
U_1&=e^{-i JD_y \tau}, \\
U_2&=e^{-i\phi Z}e^{-i JD_y \tau}e^{i\phi Z}, \\
U_3&=e^{-2i\phi Z}e^{-i JD_y \tau}e^{2i\phi Z}, \\ &\cdots \\
U_n &=e^{-i(n-1)\phi Z}e^{-i JD_y \tau}e^{i(n-1)\phi Z}
\end{align*}
where \(Z=\sum_j S_z^j\). The total evolution operator over \(n\) cycles is given by the product:
\begin{align*}
U(n \tau)&=U_n U_{n-1}\cdots U_3 U_2 U_1\\
&=e^{-i n\phi Z}\left[e^{i \phi Z} e^{-i JD_y \tau} \right]^n \\
&\approx e^{-i n\phi Z}e^{ -i \left(JD_y -\frac{\phi}{\tau}\right)n\tau}=e^{-i n\phi Z}e^{-i JD_y \tau},
\end{align*}
where the Floquet sequence is then given by \(H_1=JD_y\quad H_2=hZ\), with \(h=-\phi/\tau\).
The rotation approach also generates an extra term \(e^{-i n \phi Z}\), this term can be canceled in MQC experiments by rotating the observable by \(n\phi\).
%This approach allows us to independently tune the disordered field by adjusting \(g\), and the uniform field by varying \(b\).

We note that our methods can be applied more broadly to engineer desired Hamiltonians $ H_{des}$ using only collective rotations of the spins applied to the naturally occurring Hamiltonian, $H_{nat}$. The engineered Hamiltonian is obtained by piece-wise constant evolution under-rotated versions of the natural Hamiltonian under the condition
\(
\sum_k R_k H_{nat} R_k^\dag = H_{des},
\)
where $R_k$ are collective rotations of all the spins, which achieves the desired operator to first order in a Magnus expansion. Symmetrization of the sequence can further cancel out the lowest order correction.
Using only collective pulses limits which Hamiltonians can be engineered, due to symmetries of the natural Hamiltonian and the action of collective operators. For typical two-body interactions of spin-1/2, an efficient tool to predict which Hamiltonians are accessible is to use spherical tensors~\cite{Ajoy13l}.

\section{Fit of heating rates}
\label{app:fit}
%todo:SM (fit)
%todo: figure fitting
\begin{figure}[t]
\centering
\includegraphics[width=0.95\columnwidth]{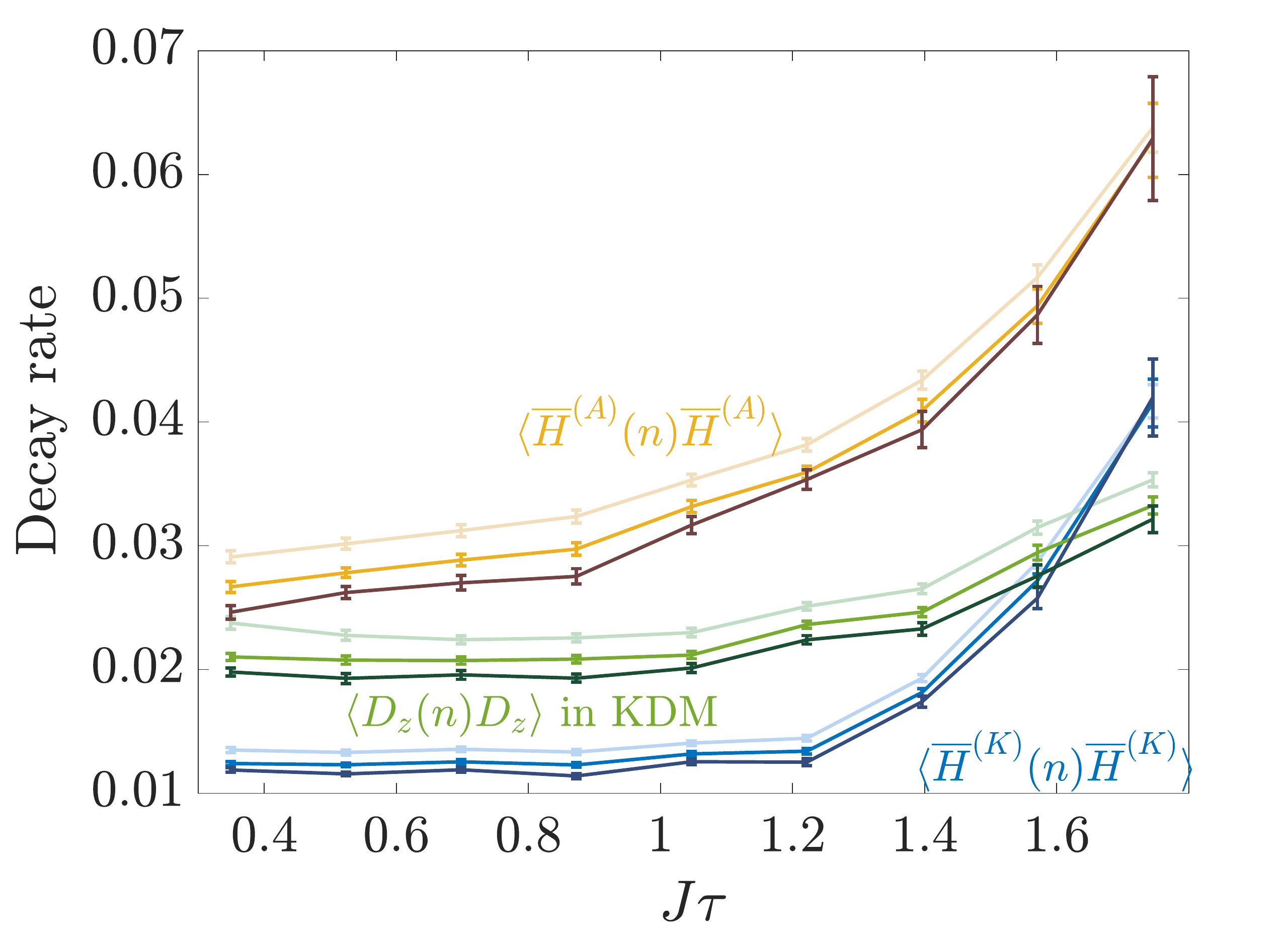}
\caption{\label{fig:DR_start}
Fitted decay rates for different fitting range: $n\in[10,64]$ (light colors), $n\in[20,64]$ (intermediate colors) and $n\in[30,64]$ (dark colors).
}
\end{figure}
\begin{figure}[t!]
\centering
\includegraphics[width=0.95\columnwidth]{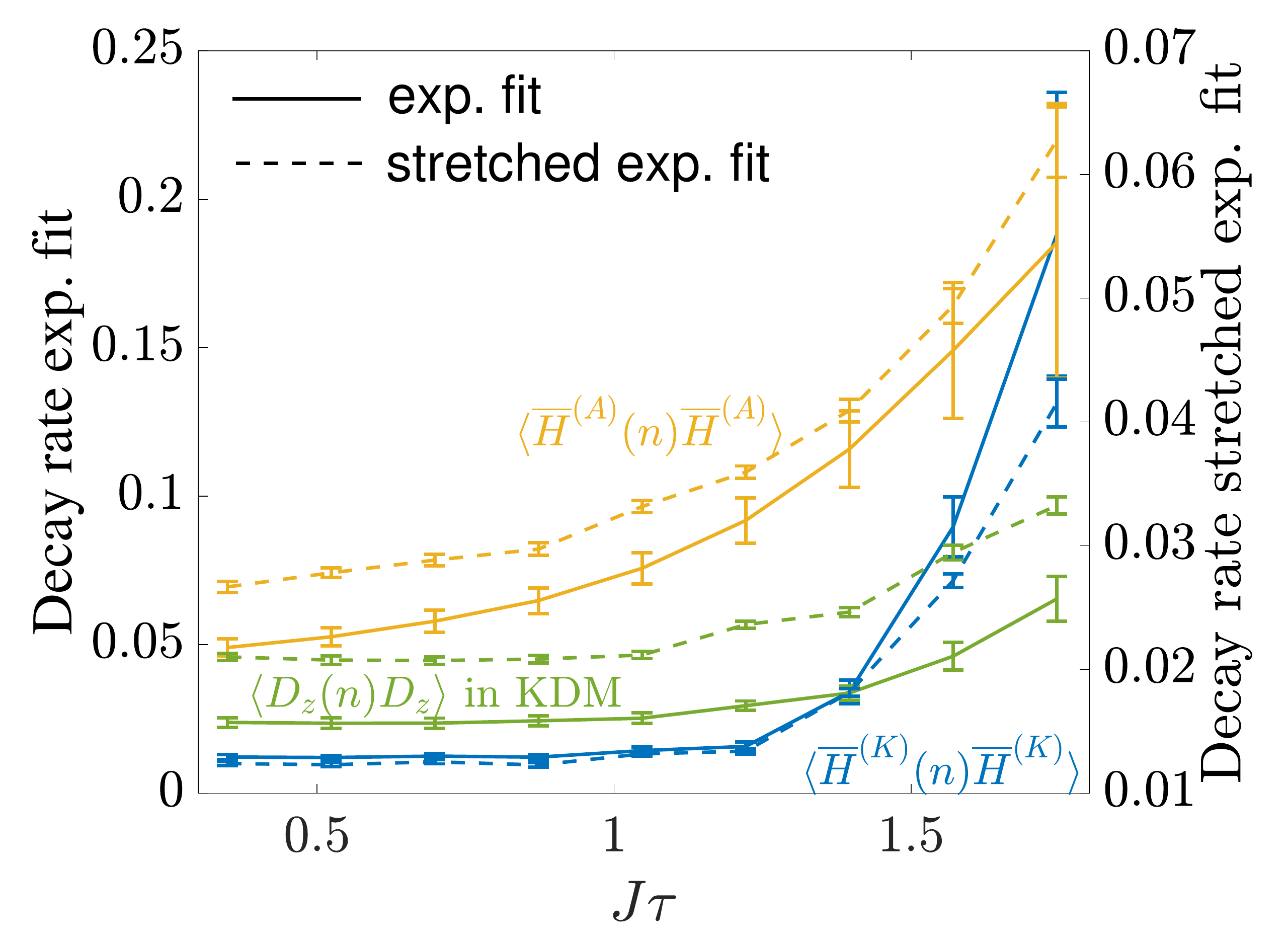}
\caption{\label{fig:DR_str_exp}
Fitted decay rates using exponential fitting (solid curves, left axis) and stretched exponential fitting (dashed curves, right axis). Fitting range is $n\in[20,64]$.
}
\end{figure}

In the main text we fit the autocorrelation decay rates by an exponential function $\langle \mathcal{O}(n)\mathcal{O}\rangle \propto \exp(-\gamma n)$ using data within range $n\in [20,64]$ to exclude transient effects. Since neither the specific form of the decay function nor the end of the transient dynamics is known exactly, in this section we present the decay rates obtained when varying the fitting range and the fitting function and show that these do not qualitatively change their behavior and thus our conclusions. Figure~\ref{fig:DR_start} depicts the decay rates obtained from an exponential fitting of the data over different ranges.
Fitting a smaller range starting at later time results in a slightly smaller decay rate (with larger uncertainty), but the exponential trend is unchanged.
In Fig.~\ref{fig:DR_str_exp} we compared fitting to an exponential and to a stretched exponential function, $\langle \mathcal{O}(n)\mathcal{O}\rangle=\langle \mathcal{O}(0)\mathcal{O}\rangle\exp(-(t/\tau_K)^\alpha)$. We choose the stretched exponential because it is a good model for exponential decays under a distribution of decay rates. Here we plot the inverse of the mean relaxation time of a stretched exponential function,
\begin{equation}
 1/\gamma=\langle\tau_R\rangle=\frac{\tau_K}{\alpha}\Gamma\left(\frac{1}{\alpha}\right),
\end{equation}
where $\Gamma$ is the Gamma function.
%and we define the decay rate $\gamma$ as $1/\langle\tau_R\rangle$.
The decay rates from both fitting models are qualitatively the same.

\end{document}